# Real-time Continuous Uncertainty Annotation for

# Spatial Navigation Studies

Qi Yang[a]

Saleh Kalantari[a, *]

[a] *Human Centered Design, Cornell University, Ithaca, NY.*

[*] Corresponding author. E-mail address: sk3268@cornell.edu

## Abstract

This study introduces two methods for continuously measuring uncertainty during human navigation in complex buildings: one using a joystick (RCUA), and the other with annotations on videos of recent navigation activity (CUA). To evaluate the usability, reliability, and validity of both approaches, we conducted a study with 54 participants. We assessed the measures' reactivity during different sign-seeing events. We also evaluated the convergent validity of both measures by comparing their outcomes with a self-report questionnaire, and assessed their discriminative and predictive validity by comparing uncertain values between known groups and correlating those values with wayfinding performance. Our findings suggest that both approaches were valid at the task level, but RCUA was better at capturing fine-grained dynamics of human experience. These continuous uncertainty measures can provide valuable insights into the fleeting nature of human experience and help identify "problem spots" for wayfinding in complex buildings.

*Keywords:* Psychological Uncertainty, Wayfinding, Continuous Assessment, Continuous Annotation, Uncertainty Assessment.

**Introduction**

Spatial navigation research is gaining importance across various fields, such as psychology, architecture, and information science due to two key reasons. Firstly, global urbanization is changing the environments people navigate in their daily lives with more complex buildings being created, which poses a challenge for humans to navigate (United Nations, 2018). Secondly, the aging population is associated with decreased ability of spatial navigation, which is a concern exacerbated by conditions such as dementia (Head & Isom, 2010; Lithfous et al., 2013; Moffat, 2009). Hence, it is crucial to understand human spatial navigation variability to develop better buildings, assistive technologies, rehabilitative strategies, and diagnostic tools to improve navigational performance.

Wayfinding is the process of navigating via a known path or finding a path to a desired destination (Wiener et al., 2009). Though wayfinding process is a continuous interactive process between human and the environment (Raubal, 2001), most existing metrics of wayfinding behaviors are *discrete* in the sense that they measure outcomes at the wayfinding task level. For example, previous researchers have measured the completion time of wayfinding tasks (O'Neill, 1992; Kuliga et al., 2019), distance traveled (Witmer et al., 1996; Hölscher et al., 2006; Gath-Morad et al., 2021), and the number of mistakes made during path selections (Dong et al., 2021). Other approaches for studying wayfinding cognition included evaluating participants' success in directional pointing tasks (Schinazi et al., 2013), the ability to remember landmark sequences (Rounds, Cruz-Garza, & Kalantari, 2020), and drawing trajectories (Rovine & Weisman, 1989). Such approaches are useful, but they can't capture the dynamics during the continuous process of wayfinding. Therefore, continuous measures need to be developed to study fine-grained wayfinding behaviors.

In recent years wayfinding researchers are increasingly using novel tools such as eye-tracking (Hollander et al., 2019), electrodermal activity (EDA) (Armougum et al., 2019), and electroencephalography (EEG) (Kalantari et al., 2022; Kalantari et al., 2023) to study continuous human perception, affective responses, and cognition. These methods require precise linkage to behavioral states, which has yet to be achieved with continuous measurement of wayfinding processes. The current research addresses this gap by developing a *continuous* metric for annotating uncertainty experience during wayfinding, allowing for studies with detailed temporal resolution of the complex interactions among environmental factors, human cognition, behavior, and wayfinding performance.

**Uncertainty during Navigation**

Psychological Uncertainty refers to a feeling of lacking information and control about incoming events (Anderson et al., 2019; Bar-Anan et al., 2009), or perceives conflicting information that challenges the person's beliefs (Hirsh et al., 2012; Jonietz & Kiefer, 2017). The Entropy Model of Uncertainty (EMU), developed by Hirsh, Mar, and Peterson (2012), suggests that if a person's internal entropy levels increase due to an inability of handling environmental challenges, alternative cognitive structures will be adopted or developed. If such structures cannot be found, the system may become overwhelmed and start to deteriorate. This entropy model is grounded in the theory that an important function of the nervous system is to minimize entropy and unpredictability by continually modifying neural structures in response to environmental information that arises during goal pursuit (Friston, 2010; Friston et al., 2006). The EMU model describes two primary domains of Uncertainty: (a) perceptual uncertainty related to sensory input, and (b) behavioral uncertainty related to the selection of action/behavior (Hirsh et al., 2012). Both of these domains are relevant to wayfinding research.

Perceptual Uncertainty often arises during wayfinding as a result of obscured, unreliable, or conflicting information sources in the environment (Brunyé et al., 2015; Martin & Richter, 2015; Maruyama et al., 2017). Behavioral wayfinding Uncertainty can be a downstream outcome of such perceptual issues, but it may also be affected by other factors such as the number of possible routes and competing goals. Uncertainty states in wayfinding are linked to behaviors such as heightened information-seeking (Keller et al., 2020) and more risk-averse wayfinding strategies (Martin & Richter, 2015). Anderson and colleagues (Anderson et al., 2019) presented that experiences of Uncertainty can also be linked to stress, anxiety, and feelings of insecurity, which may affect the long-term mental health. Zhu and colleagues (Zhu et al., 2022) found that behavioral uncertainty during human wayfinding has neurophysiological correlates which can be used to classify such uncertainty events. However, despite the importance of Uncertainty states in wayfinding, there is currently no validated tool to measure such states continuously.

**Continuous Self-reported Measurements in Behavioral Studies**

Compared to discrete post-stimulus measurements, continuous evaluation offers advantages. When the data are self-reported, continuous reporting is generally less intrusive and less likely to suffer from response biases or to interrupt the participants' experiences of the phenomena being studied (Wagner et al., 2021). Discrete measurements are most useful for short-duration stimuli that do not unfold through time, such as a still image (Fayn et al., 2022; Girard & Aidan, 2018). In contrast, continuous measurements can provide data about changing human responses during ongoing behavioral processes, as is the case for wayfinding activities. The approach that we developed for measuring wayfinding Uncertainty was loosely based on continuous self-report measures that have been used in other types of behavioral research such as studies of music reception, affective computing, and interpersonal interactions (Biocca et al., 1994; Cowie et al., 2000; Fuentes et al., 2017; Gabriel et al., 2016; Geringer et al., 2004; Lizdek et al., 2012; Madsen, 1997; Ruef & Levenson, 2007; Sadler et al., 2009; Schubert, 2010). The common approach in these studies was to ask participants to engage in a continuous task, such as watching a video or engaging in a conversation, while simultaneously using some type of hand-held controller to continuously report the variable of interest.

One issue in such approaches is that the continuous response reporting may divide the participants' attention by increasing cognitive or physical load (Li et al., 2015; Malandrakis et al., 2011; Yannakakis & Martínez, 2015; Kalantari et al., 2021; Zhang et al., 2020). This problem was most notable when using physical devices such as mouse that required users to make constant adjustments. For this reason, researchers have gravitated toward the use of joysticks that include a return spring and thus automatically realign to the center in the absence of force (Li et al., 2015; Girard, 2014; Metallinou & Narayanan, 2013; Nicolaou et al., 2014). This creates a more intuitive response-interface in which the user does not need to devote any specific cognitive attention to evaluating the position of the controller, but only needs to be aware of the amount of force applied. We adopted this approach by using a joystick device for our Real-time Continuous Uncertainty Annotation (RCUA). Moreover, as the older adult is one increasing population of wayfinding research (Bosch & Gharaveis, 2017). We investigate if older adults have challenges in using RCUA.

An additional concern in wayfinding tasks is that participants should not be visually distracted by the response interface, as they will need to have their full visual focus available to perceive the environment in a natural fashion. (This is not necessary when participants are, for example, listening to music or engaging in conversations.) In most of the prior studies that involves continuous self-reported measures, participants were shown the values that they were reporting in real-time on a display screen, which allowed them to make adjustments based on that visual feedback. The use of a joystick in natural environments eliminates the feedback display, but this introduces potential concerns about response accuracy (Xue et al., 2021). To address this concern and evaluate the impact of eliminating visual feedback, we developed a second type of Continuous Uncertainty Annotation (CUA) that was not conducted in real-time. For the CUA measure, participants were first asked to complete wayfinding tasks, and then they watched the first-person video of their endeavor, while using a physical slider to annotate how uncertain they felt during the experience. In this approach the participants were susceptible to greater memory bias but they could see the exact value of Uncertainty that they were reporting displayed on the screen.

**Research Aims**

    **Validity of RCUA and CUA.** We studied the continuous measures' discriminative validity at the wayfinding task level (H1) and the fine-grained sign-seeing level (H2) by using known groups (Cronbach & Meehl, 1955). We examined whether the metrics predicted wayfinding performance (H3). To establish convergent validity (Campbell & Fiske, 1959), we compared the uncertainty metrics derived from the RCUA and the CUA with a self-report uncertainty questionnaire (H4). Finally, we assessed the similarity between RCUA and CUA metrics (H5). Therefore, we hypothesized that:

    H1: Novice wayfinders will report higher Uncertainty for wayfinding tasks in comparison to Trained wayfinders and Expert wayfinders using the RCUA and the CUA.

    H2: Participants will report less Uncertainty after seeing a sign with directional guidance of the task, while they will report the same or higher Uncertainty after seeing signs without directional guidance using both RCUA and CUA.

H3: Higher Uncertainty levels as reported by the RCUA and the CUA will be associated with longer wayfinding task Completion Times and greater Distance Traveled.

H4: Participants' discrete Uncertainty levels on a self-report questionnaire will be correlated with their RCUA and CUA measurements.

H5: Responses on the RCUA and the CUA for the same wayfinding task will have no significant differences for similar Uncertainty metrics.

**Reliability of RCUA and CUA.** To understand how well participants are able to consistently report desired uncertainty level using RCUA. We asked participants to repeatedly report among four levels of Uncertainty by exerting different amounts of pressure on the joystick under three mobile conditions. We hypothesized that:

H6: There will be significant differences in the RCUA values between the four levels of Uncertainty ("None," "Slight," "Moderate," and "Extreme"), during all three conditions.

**Usability of RCUA.** One of our concerns was to determine if participants would be able to successfully use the RCUA joystick to report Uncertainty levels during an active wayfinding task, and if they could do so without excessive distraction from the wayfinding task. To this end, we asked participants to self-report their perceived cognitive load and how often they had forgotten to use the joystick after the wayfinding tasks.

## Methods

This reliability, validity, and usability study consisted of three main steps. First, we test the RCUA's reliability by measuring how consistently the participants were able to report different levels of Uncertainty using the joystick device under a variety of conditions. Second, we asked the participants to use the device to continuously report their Uncertainty levels while completing several wayfinding tasks in a building interior. This allowed us to evaluate the construct validity, discriminative validity, and predictive validity of the RCUA. Finally, we asked the participants to review a first-person video of their wayfinding activities immediately after completing them, during which time they provided the CUA annotation. This allowed us to evaluate the validity of the CUA and compare its results against the RCUA. We explain the details in the Procedure section.

**Procedures**

Prior to the experiment sessions, participants were asked to complete an online survey that collected demographic information along with responses for the SAS and the SBSOD. When participants arrived in the lab for their sessions, the researchers administered the MoCA and the Mental Rotation Test. After completing these enrollment surveys, the participants were asked to finish Parts 1–3 of the experiment as described below.

**Part 1: Training.** In this portion of the experiment, which we called the "Repeated Push," participants were instructed to use the joystick to report four levels of uncertainty ("None," "Slight," "Moderate," and "Extreme"). First, they were allowed 5 to 15 minutes of practice with a computer-based training program, which provided visual feedback regarding the joystick push level (as both a progress bar and a number). After the training, the researcher randomly asked the participants to report a particular level of uncertainty without visual feedback. This was repeated 8 times (twice for each uncertainty level), for each of three conditions: (a) seated in the lab before any wayfinding tasks, (b) walking through the lab before any wayfinding tasks, and later (c) walking through the lab after completing all wayfinding tasks. The purpose of this activity was to determine if the participants could accurately use the joystick to report the intended uncertainty level.

**Part 2: RCUA Annotation.** All participants, including Novice, Trained, and Expert wayfinders, were asked to complete the same wayfinding tasks in the same two interconnected buildings. The site of these activities was the **[deleted for the purpose of blind review]**, which are connected through a commons area. The connection of an old building that has undergone multiple rounds of renovation with a new building make this environment very complex for wayfinding. **[Deleted for the purpose of blind review]** has received numerous complaints about the difficulties of finding destinations in this environment, making it ideal for studying uncertainty in indoor navigation. The study participants were asked to find seven specific destinations within the buildings in a set sequence (Figure 3 and supplementary information s1.). During each wayfinding task, the participants used the RCUA to continuously report their wayfinding Uncertainty level. After participants complete each wayfinding task, they were asked to report their overall perceived Uncertainty level of the task using a self-report questionnaire, and report their task load. They were then informed of their next wayfinding goal. After finishing all the tasks, the participants and researchers returned to the lab, conduct CUA annotation and self-report the level of wayfinding interference using the RCUA.

Part 3: CUA Annotation. In the lab-based annotation, participants relied on their memory of the recent wayfinding tasks and used the CUA slider to indicate levels of Uncertainty while watching first-person video taken from their body-cam. The current value of the CUA slider was displayed at the top-right corner of the screen during the video, so that the participants could observe their input. To reduce the length of the experiment sessions and avoid participant fatigue, and potentially to help reduce response bias, the video was played at 2x speed while participants made these Uncertainty annotations.

**Special Procedure for "Trained" Wayfinders.** The "Trained" participant group was asked to undertake a somewhat different and more arduous procedure. This group participated in Parts 1, 2, and 3 of the experiment as Novices, as described above. After doing so, however, they were asked to return to the on-site buildings and complete Parts 2 and 3 again, after looking at a printed map with all destinations and shortest paths highlighted. Reviewing the map served as a known treatment, as this has been shown to improve wayfinding performance in multiple studies (Hölscher et al., 2007; Link et al., 2011). After examining the map, the Trained participants conducted all seven of the wayfinding tasks a second time with RCUA reporting, and then they returned to the lab and completed the CUA reporting a second time. A full summary of the experiment procedure is illustrated in Figure 1.

**Part4: Sign Seeing Events Annotation.** After the experiment, one researcher watched the first-person video recordings of the wayfinding process and annotated the sign seeing events. The starting point is marked when the participant stops and starts to read the sign. If the sign contains directional guidance of the task, it's marked as helpful. Since we found most participants finished reading the signs in ten seconds. We extracted the RCUA and CUA values of ten seconds before and after the sign seeing events and excluded the events in which participants report zero uncertainty all the time. In total, 110 helpful sign seeing events and 185 unhelpful sign seeing events were extracted.

**Figure 1**

*Experimental Procedure*

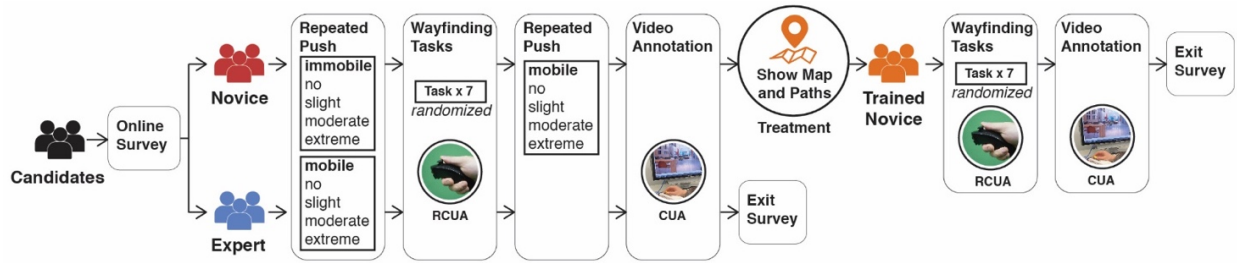

**Sample Size and Statistical Power**

An a-priori power analysis of the required sample size was conducted using G*Power software for hypothesis one (Faul et al., 2007). In order for comparison between Novice and Expert wayfinders, we need to assume mean and standard deviation for different groups. Though we were unable to find any prior literature on wayfinding Uncertainty that would provide assumptions for mean and standard deviation. we grounded our assumptions on Hölscher (2006), who studied the frequency of becoming lost during wayfinding comparing novices ($M = 0.42$, $SD = 0.17$) vs. experienced participants ($M = 0.17$, $SD = 0.21$). Since we used seven wayfinding tasks for each participant, the inter-class correlation was assumed to be n = 7, phi = 0.5, and the design effect was assumed to be 1.3. The analysis suggested that we would need a minimum of 10 participants in each group to achieve a power of 0.8 for with an alpha of 0.05. Regarding the comparison between Novice and Trained wayfinders, we assumed a large effect size (d = 0.9) and the analysis suggested 13 participants to achieve a power of 0.8 for matched-pair analysis with an alpha of 0.05.

For other hypotheses (H2, H5, H6), we didn't conduct prior power evaluation. We conducted post-hoc simulation-based sensitivity analysis using simr package in R (Green &

Macleod, 2016) to evaluate the minimum effect size that our sample size will be able to detect with a power of 0.8.

**Participants**

Participants eligible for the study had to meet the following criteria: the ability to successfully perform wayfinding tasks in an indoor setting, including using stairs; no significant motor impairments; proficiency in understanding written and spoken English; and a score of 18 or above on the Montreal Cognitive Assessment (MoCA). (Nasreddine et al., 2005) as administered by the researchers. (The MoCA was used to screen out participants who had possible cognitive impairments.) A total of 69 participants were recruited for the study using a convenience sampling method. Data from 15 of the participants had to be excluded due to technical issues, including several instances of overheated or crashed computers during the experiment sessions, one instance of a disconnected wireless adapter, and several instances of mistakes by the research team. The researcher mistakes included forgetting to initiate data-recording and accidentally deleting raw data during transfer. This was a higher than anticipated attrition rate, but the remaining 54 participants were sufficient to conduct our statistical analysis.

The final sample included 11 "Expert" wayfinders, designated as such because they were closely familiar with the two interconnected buildings that were used in the study (having visited the buildings a minimum of ten times and expressing confidence in giving wayfinding directions for the buildings). The remaining 43 participants had previously visited the buildings no more than once, and were designated as "Novice" wayfinders. From the pool of novices, 13 were randomly selected to become "Trained" wayfinders. This select group completed the wayfinding tasks twice, and before their second, "trained" attempt, they were allowed to review a floorplan-map of the buildings, on which the optimal routes for the study's wayfinding tasks were highlighted. (Thus, the "Trained" group provided data twice, once as part of the Novice group and then a second time as Trained.)

Given the importance of age-related factors in wayfinding cognition (Head & Isom, 2010; Lithfous et al., 2013; Moffat, 2009), we placed an emphasis on recruiting participants from across the adult human lifespan. This was done by reaching out to **[removed for the purpose of blind review]** as part of the recruitment process. The 54 participants who were included in the study ranged in their reported ages from 18 to 77 years old ($M = 33.4$, $SD = 20.9$), including fifteen participants (28%) who indicated that they were 60 years or older.

When asked about their gender, 36 participants reported as Female, 17 as Male, and 1 as Other. Detailed distribution of gender of each group is shown in Table 1.

All participants gave informed written consent prior to the research activities, and the study procedures were reviewed and approved by the Institutional Review Board at **[removed for the purpose of blind review]**.

**Table 1**

*Gender Distribution of Each Group. N: Novice, E: Expert, T: Trained.*

|  | Female | Male | Other |
| --- | --- | --- | --- |
| Old | N = 9 | N = 5 | N = 0 |
|  | E = 1 | E = 0 | E = 0 |
|  | T = 0 | T = 0 | T = 0 |
| Young | N = 19 | N = 9 | N = 1 |
|  | E = 7 | E = 3 | E = 0 |
|  | T = 9 | T = 4 | T = 0 |

**Hardware and Software Components of RCUA and CUA**

The components of the RCUA and CUA were selected in such a way that they could easily be assembled and would be accessible to other researchers. For the RCUA joystick, we used a Nintendo Wii Nunchuk Controller paired with an 8BitDo Wireless USB Adapter (Figure 2a). The controller was chosen because it has a large analog stick, which is convenient for users to hold and push. In line with recent studies on human motor (Loram et al., 2011), we developed a script in the Python language that documented the real-time input of the controller in terms of the amount of force exerted in any direction from center. This allowed us to record a very simple and precise continuous metric of the user's reported Uncertainty level, which we placed on a scale of 0 (no force / no Uncertainty) to 1 (maximum force / maximum Uncertainty). Participants were trained in the use of the joystick for reporting Uncertainty levels prior to the wayfinding tasks, as discussed in the Procedures section below.

During the real-world wayfinding tasks, one or two researchers followed the participants at approximately 5–10 ft. distance, carrying a tablet computer (Microsoft Surface Pro). This computer received and recorded the wireless data from the joystick. Simultaneously, it ran the **[removed for the purpose of blind review]** App previously developed by our team, which documented the participants' physical trajectories through the building over time. To create video of the wayfinding process, both the participant and the following researcher wore body cameras (GoPro Max), positioned at chest level. The participant's first-person video was used later in the CUA portion of the study, while the following researcher's third-person video was not used in the current analysis.

For the CUA metric we used a physical slider device, specifically the linear slide potentiometer module analog sensor. This selection was based on relevant prior work in video annotation (Fayn et al, 2022; Fuentes et al., 2017), which has indicated that such a tangible device providing proprioceptive feedback is more effective for collecting participant reactions compared to mouse input. The slide-sensor was connected to an Arduino Uno board, which was contained within a white casing and linked to a standard desktop computer via USB-A (Figure 2b). We developed a processing script in Python that played the wayfinding video while documenting and synchronizing the real-time input of the slide. Similar to the RCUA, the slide input for the CUA was scaled from 0 (no Uncertainty) to 1 (maximum Uncertainty). The videos of the participants' prior wayfinding activities were presented on flat-screen displays with a resolution of 1920 x 1080 and a 60 Hz. refresh rate. Both RCUA and CUA data were down sampled to 5 Hz linearly. All of our software components have been made freely available for use by other researchers.

**Figure 2**

*(A) Experiment Setting for Real Time Uncertainty Annotation (RCUA) during Wayfinding Tasks, and (B) Experiment Setting for Continuous Uncertainty Annotation (CUA) after the Wayfinding Tasks*

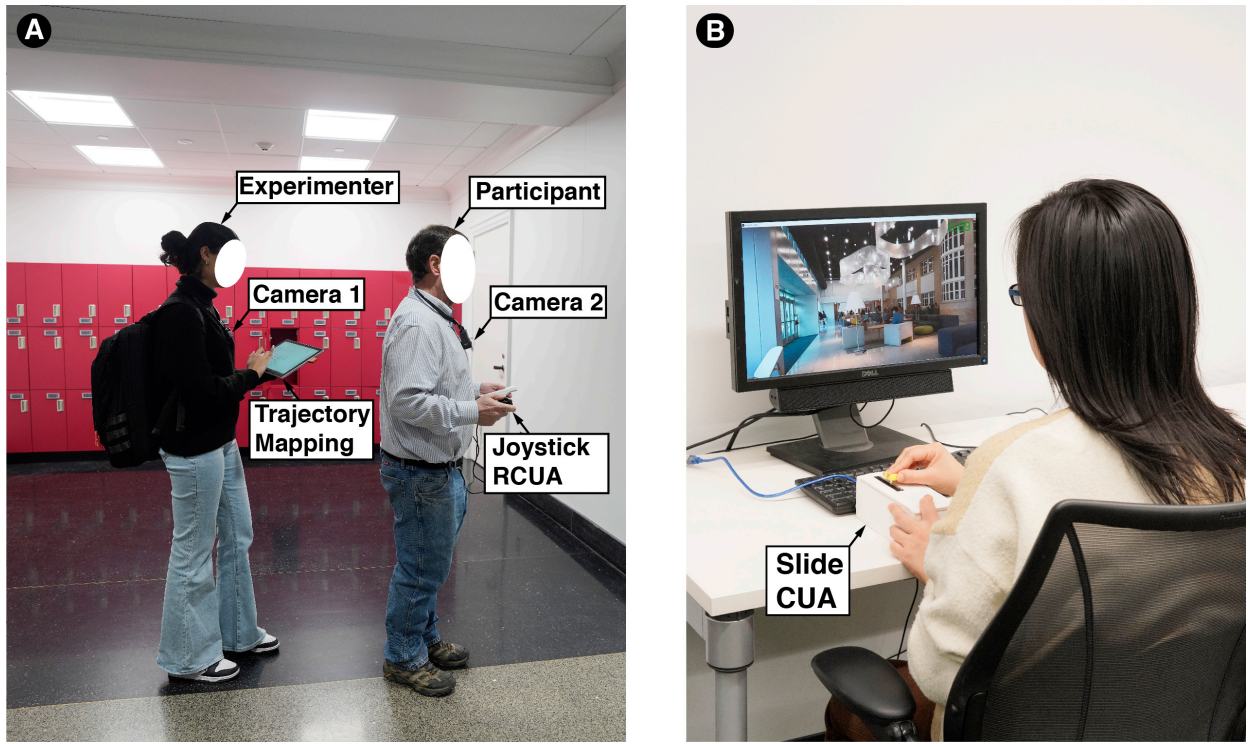

**Measures**

*Continuous Uncertainty*. The continuous Uncertainty data recorded via both RCUA and CUA were operationalized into four levels: "No Uncertainty" (RCUA/CUA = 0), "Slightly Uncertain" (0 < RCUA/CUA ≤ 0.3), "Moderately Uncertain" (0.3 < RCUA/CUA < 1) and "Extremely Uncertain" (RCUA/CUA=1). The data across an entire wayfinding task for one participant were summarized in five metrics: the total sum of reported uncertainty ($S_U$), the average uncertainty level ($M_U$), the total duration of uncertainty ($T_U$), the total duration of "Extreme" uncertainty ($T_{EU}$), and the frequency of uncertainty ($F_U$) (Formulas are shown in supplementary information s2.).

*Self-report Task Uncertainty.* The discrete task uncertainty questionnaire is adapted from the one used in the previous study (Kalantari et al., 2022) with 10 levels (1= No Uncertainty, 10 = High Uncertainty).

*Task Load.* We use NASA-TLX Questionnaire to measure task load (Hart & Staveland, 1988), which measures workload across six dimensions—mental demand, physical demand, temporal demand, effort, performance, and frustration—each has 10 levels, with higher scores indicating greater perceived task loads.

*Cognitive Impairment.* As noted above in the participant inclusion criteria, we used the Montreal Cognitive Assessment to screen for possible cognitive impairment, a factor that is particularly relevant for older adults. Scores of 17 or lower on the MoCA are regarded as an indicator of possible moderate or severe cognitive impairment and were excluded (Carson et al., 2018; Milani et al., 2018; Wong et al., 2015).

*Spatial Anxiety.* The Spatial Anxiety Scale (SAS) (Lawton, 1994) was used to assesses participants' feelings of apprehension related to everyday tasks requiring spatial or navigational skills. This is an 8-item, 5-point Likert scale instrument in which higher scores indicate that the respondent associates greater anxiety with spatial tasks.

*Spatial Abilities.* Spatial abilities were also measured using the Santa Barbara Sense of Direction Scale (SBSOD) (Hegarty, Mary, Anthony E. Richardson, Daniel R. Montello, Kristin Lovelace, 2005). This is 15-item, 7-point Likert scale instrument in which higher scores indicate greater difficulty in spatial orientation, a factor that has previously been associated with wayfinding performance (Kuliga et al., 2019). We also asked participants to complete the Mental Rotation Test (Vandenberg & Kuse, 1978), another metric commonly used to measure spatial ability. This instrument asks participants to mentally visualize rotating a three-dimensional object and to select the correct image from a list of options. Higher scores on this measure are interpreted as indicating greater spatial abilities.

*Self-report Wayfinding Interference.* We measure self-report wayfinding interference by asking the participants to rate the extent to which using the RCUA interfered with their wayfinding performance (1 = No Inference, 10 = High Inference). In addition, we asked participants to estimate the number of times they forgot to use the RCUA joystick to report uncertainty.

**Figure 3**

*Map of the Seven Wayfinding Tasks, Showing the Physical Locations of Highest Uncertainty*

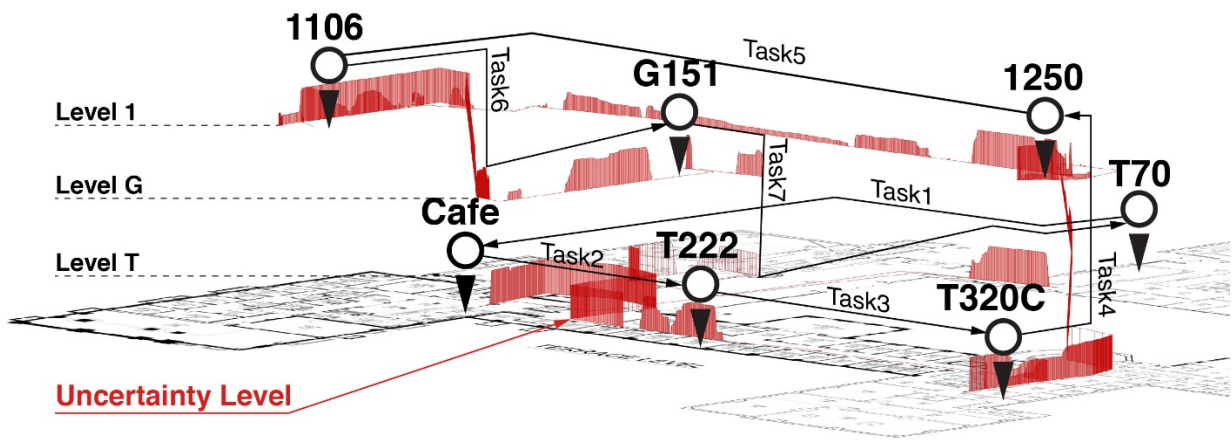

**Statistical Analysis**

For discriminative validity (H1), we determined if there was a difference between novice and trained wayfinders by using a linear-mixed model in which participants experience, participant number, task, and the interaction between participant and task were used as random effects. Since comparison between novice and expert was a between-group comparison, besides random effects mentioned before, we added random effects including sense of direction and spatial anxiety. To test if uncertainty drops after seeing signs and if RCUA and CUA has different differences (H2), we averaged the uncertainty values before and after seeing signs and fitted a linear-mixed model in which annotation methods and time were used as fixed effect. Regarding predictive validity (H3), we firstly calculated the z-scores of the Uncertainty measures within each participant, and fitted a linear-mixed model to evaluate the intra-class correlation between proposed measures and wayfinding performance metrics separately for each task, and determine the average. For convergent validity (H4), we used the similar method in H3 to evaluate the intra-class correlation between proposed measures and self-report Uncertainty questionnaire measure. To test if RCUA and CUA provided different metrics for the tasks (H5), we fitted linear-mixed models to evaluate the significance of the fix effect of the measure on Uncertainty metrics. To analyze the RCUA reliability (H6), we fitted linear-mixed models to evaluate if four levels of uncertainty were significantly different from each other. We included participant number as random effect, and included fixed effect of desired uncertainty levels (no, slightly, moderate, extreme), and condition (pre-immobile, pre-mobile, post-mobile), and interaction. We repeated this model separately for younger and older adults.

We reported descriptive statistics of the NASA-TLX, self-report interference and number of forgets to demonstrate the feasibility and usability of the RCUA. To measure how RCUA matches CUA at the temporal level.

Since we were testing significance for five metrics ($S_U$, $M_U$, $T_U$, $T_{EU}$, and $F_U$) of RCUA and CUA in H1, we used Bonferroni correction to adjust our p-values.

# Results

## Discriminative Validity of RCUA and CUA (Hypotheses 1)

We found significant differences between the scores of Novice vs. Trained wayfinders for all five discrete RCUA and CUA metrics ($S_U$, $M_U$, $T_U$, $T_{EU}$, and $F_U$). The metrics for Average Uncertainty ($M_U$) and Total Uncertainty Duration ($T_U$) had the largest effect sizes (Figure 4 and Table 2).

**Figure 4**

*Average Uncertainty Value ($M_U$) of Each Wayfinding Task, Comparison between Novice Wayfinders and Trained Wayfinders*

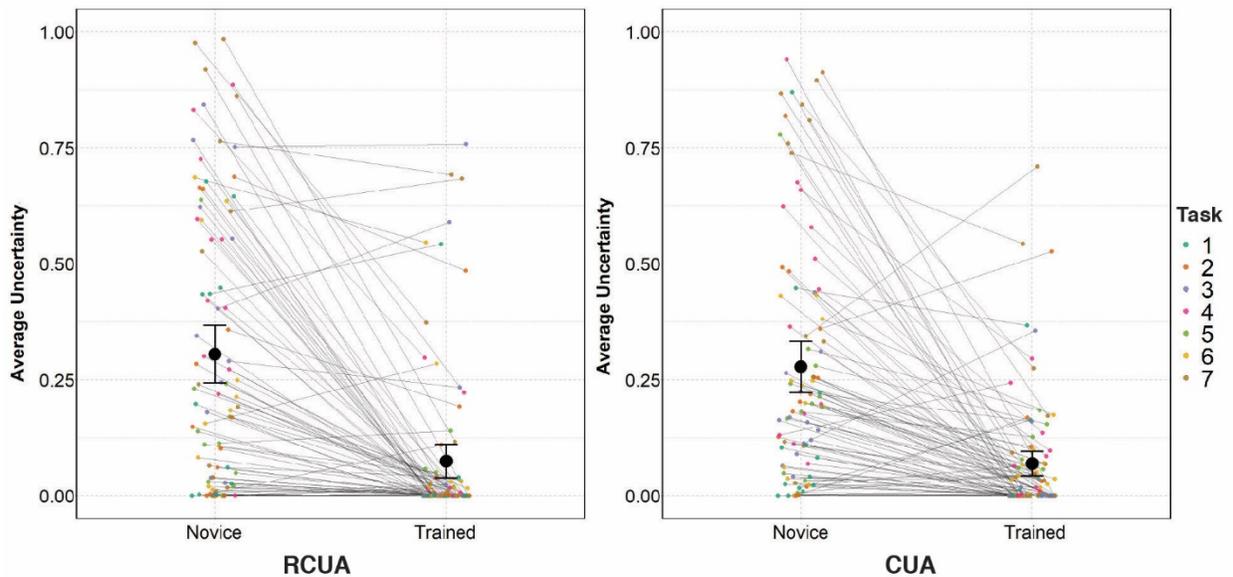

Regarding the comparison between Novice vs. Expert wayfinders, we found significant differences in the Total Uncertainty Duration ($T_U$) and the Frequency of Uncertainty ($F_U$) for the RCUA. For the CUA, we found significant differences for the Average Uncertainty ($M_U$), as well as for Total Uncertainty Duration ($T_U$) and Frequency of Uncertainty ($F_U$) (Figure 5 and Table 3). We did not observe any statistically significant differences in the spatial abilities and spatial anxiety between the "Novice" and "Expert" wayfinding groups, which means that they did not act as confounding variables in the study. These results were less decisive than the Novice vs.

Trained comparisons. In summary, H1 is partially supported.

**Table 2**

*Novice vs. Trained Comparisons for All RCUA and CUA Metrics*

|  | *t* | *d* | *p* |
|---|---|---|---|
| RCUA |  |  |  |
| Average | 7.902 | 1.67 | < 0.0001**** |
| Sum | 5.952 | 1.25 | < 0.0001**** |
| Duration | 7.314 | 1.54 | < 0.0001**** |
| Extreme | 5.413 | 1.14 | < 0.0001**** |
| Frequency | 7.606 | 1.61 | < 0.0001**** |
| CUA |  |  |  |
| Average | 7.410 | 1.56 | < 0.0001**** |
| Sum | 5.522 | 1.16 | < 0.0001**** |
| Duration | 7.852 | 1.66 | < 0.0001**** |
| Extreme | 2.940 | 0.62 | 0.0041* |
| Frequency | 7.834 | 1.25 | < 0.0001**** |

*Note.* p-threshold is 0.05 / 10 = 0.005 after Bonferroni correction. *significant after correction*

**Table 3**

*Novice vs. Expert Comparisons for All RCUA and CUA Metrics*

|  | *t* | *d* | *p* |
|---|---|---|---|
| **RCUA** | | | |
| Average | 2.739 | 0.81 | 0.0088 |
| Sum | 2.018 | 0.81 | 0.0494 |
| Duration | 3.047 | 0.91 | 0.0039* |
| Extreme | 2.457 | 0.73 | 0.0180 |
| Frequency | 2.869 | 0.80 | 0.0059 |
| **CUA** | | | |
| Average | 3.125 | 0.93 | 0.0031* |
| Sum | 1.828 | 0.54 | 0.0740 |
| Duration | 3.226 | 0.97 | 0.0024* |
| Extreme | 1.551 | 0.46 | 0.1278 |
| Frequency | 3.592 | 1.00 | 0.0007** |

*Note.* p-threshold is 0.05 / 10 = 0.005 after Bonferroni correction. * *significant after correction*

**Figure 5**

*Average Uncertainty Value (M$_U$) of Each Wayfinding Task, Comparison between Novice*

*Wayfinders and Expert Wayfinders*

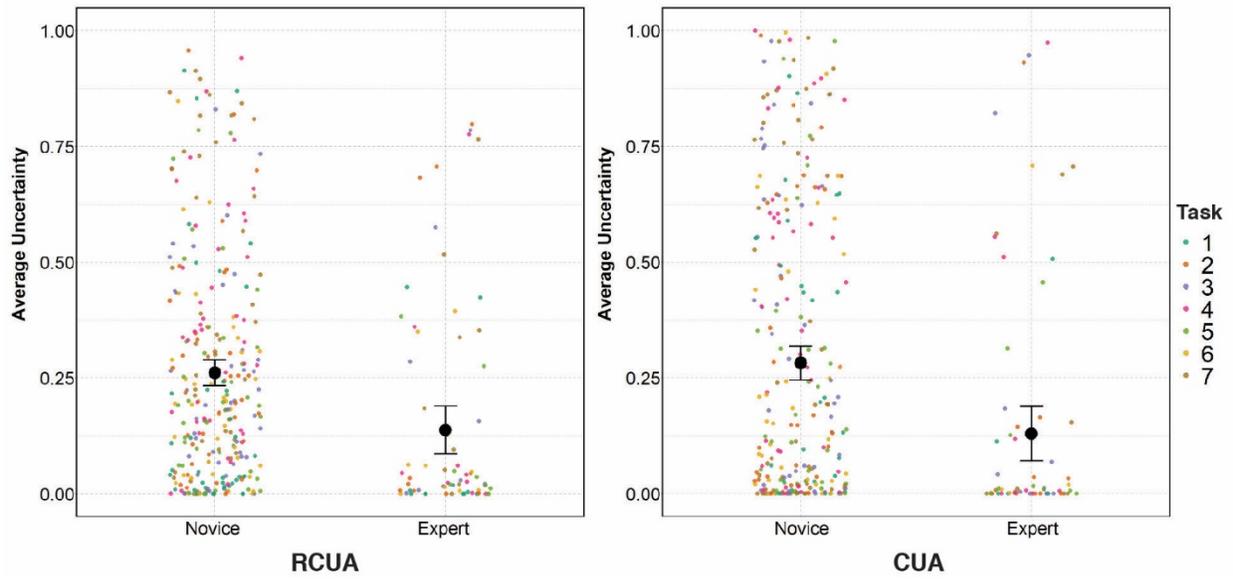

**Figure 6**

*RCUA and CUA Values Before and After Seeing Helpful vs. Unhelpful Signs*

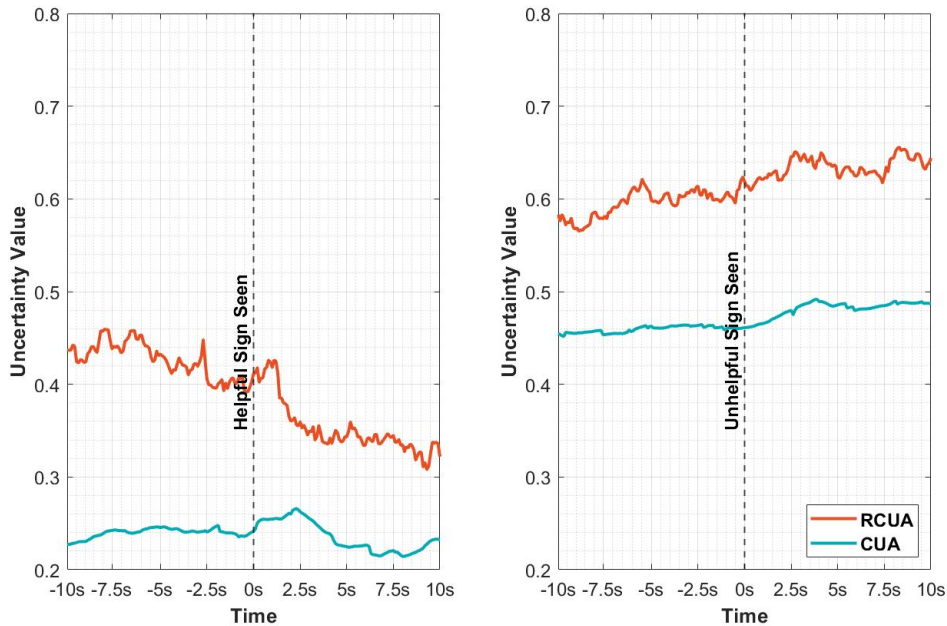

**Reactivity of RCUA and CUA Measure (Hypothesis 2)**

We observed when participants approached and started to read the helpful signs, the uncertainty level continued to drop for the RCUA. However, Uncertainty level of CUA remained relatively unchanged regardless whether the sign is helpful or unhelpful (Figure 6). We also noticed RCUA Uncertainty values are higher than CUA Uncertainty values in both sign conditions ($p < 0.05$). We found significant difference between the average RCUA uncertainty value before and after seeing a helpful sign ($t = 3.190$, *df = 218*, $d = 0.43$, $p = 0.0016$). We didn't find significant difference of CUA Uncertainty value ($p = 0.851$). The post-hoc simulation analysis showed with our sample size, we had 80% power to detect a 0.037 decrease of CUA value as we go from before to after.

As for the unhelpful sign-seeing events, we found a small significant increase of RCUA values after seeing an unhelpful sign (*t = 2.531, df = 184, d =0.37,  p = 0.012)*. We also find a significant increase of CUA Uncertainty values ($t = 2.288$, $df = 184$, $d = 0.34$, $p = 0.023$). Therefore, Hypothesis 2 is supported.

**RCUA and CUA Predictive Validity (Hypothesis 3)**

The metrics derived from both the RCUA and the CUA showed moderate to strong correlation with the two wayfinding task-performance variables, Completion Time and Distance Travelled. The Sum Uncertainty ($S_U$) and Average Uncertainty ($M_U$) metrics were the best predictors of wayfinding performance (Table 4). H6 is supported.

**Table 4**

*Intra-class Correlation Showing Relationships between Self-reported Uncertainty, Wayfinding Task Completion Time, Wayfinding Distance Travelled, and Five Metrics ($S_U$, $M_U$, $T_U$, and $T_{EU}$) Derived from the RCUA and CUA.*

| | RCUA | | | | |
| --- | --- | --- | --- | --- | --- |
| | $S_U$ | $M_U$ | $T_U$ | $T_{EU}$ | $F_U$ |
| Self-reported Uncertainty | 0.65 | 0.53 | 0.49 | 0.50 | 0.53 |
| Task completion time | 0.71 | 0.38 | 0.38 | 0.41 | 0.54 |
| Distance Travelled | 0.68 | 0.43 | 0.38 | 0.38 | 0.53 |
| | CUA | | | | |
| | $S_U$ | $M_U$ | $T_U$ | $T_{EU}$ | $F_U$ |
| Self-reported Uncertainty | 0.65 | 0.62 | 0.25 | 0.41 | 0.42 |
| Task completion time | 0.73 | 0.52 | 0.23 | 0.27 | 0.57 |
| Distance Travelled | 0.70 | 0.49 | 0.22 | 0.26 | 0.55 |

**RCUA and CUA Convergent Validity (Hypothesis 4)**

Sum ($S_U$) and Average ($M_U$) for the RCUA and the CUA had a moderate intra-class correlation with self-reported Uncertainty questionnaire (ICC > 0.5). The highest ICC were for the Sum Uncertainty measure of both the RCUA (r = 0.65) and the CUA (r = 0.65). Overall, these results indicate a high convergent validity between the results of uncertainty questionnaire results and the results of the RCUA and CUA; thus, H4 is supported.

**Comparison of RCUA vs. CUA (Hypothesis 5)**

We found significant differences in $T_U$ and $T_{\text{EU}}$ (*p < 0.001*). We did not find any significant differences between the RCUA vs. the CUA metrics regarding $S_U$, $M_U$ and $F_U$. We picked the non-significant model with the largest effect size to run the post-hoc simulation analysis. It showed with our sample size, we had 80% power to detect a 0.038 decrease of CUA $M_U$ value compared with RCUA, which is small. Therefore, H5 is partially supported.

**RCUA Reliability (Hypothesis 6)**

In the "Repeated Push" section of the experiment, we found statistically significant differences between the RCUA joystick push values for the four levels of Uncertainty reported ("None," "Slight," "Moderate," and "Extreme"), for all three conditions in which we tested these responses (pre-wayfinding seated/immobile; pre-wayfinding walking/mobile; and post-wayfinding walking/mobile) (overall, F(3, 1114) = 345.64, *p* < 0.001). Except for the Moderate–Extreme comparison under the pre-mobile condition, the confidence intervals of adjacent Uncertainty levels did not overlap (Table 5). Thus, H2 is supported.

**Table 5**

*The 95% Confidence Intervals for Four Levels of Uncertainty ("None," "Slight," "Moderate,"*

*and "Extreme") Across the Three Experimental Conditions*

|  | **Pre-immobile** | **Pre-mobile** | **Post-mobile** |
|---|---|---|---|
| None | [0.145, 0.270] | [0.125, 0.250] | [0.164, 0.290] |
| Slight | [0.407, 0.516] | [0.496, 0.604] | [0.444, 0.553] |
| Moderate | [0.767, 0.875] | [0.810, 0.918] | [0.761, 0.871] |
| Extreme | [0.897, 1.022] | [0.892, 1.017] | [0.908, 1.034] |

**Figure 7**

*The 95% Confidence Intervals of Peak Joystick Value when Reporting Four Levels of*

*Uncertainty under Three Conditions:*

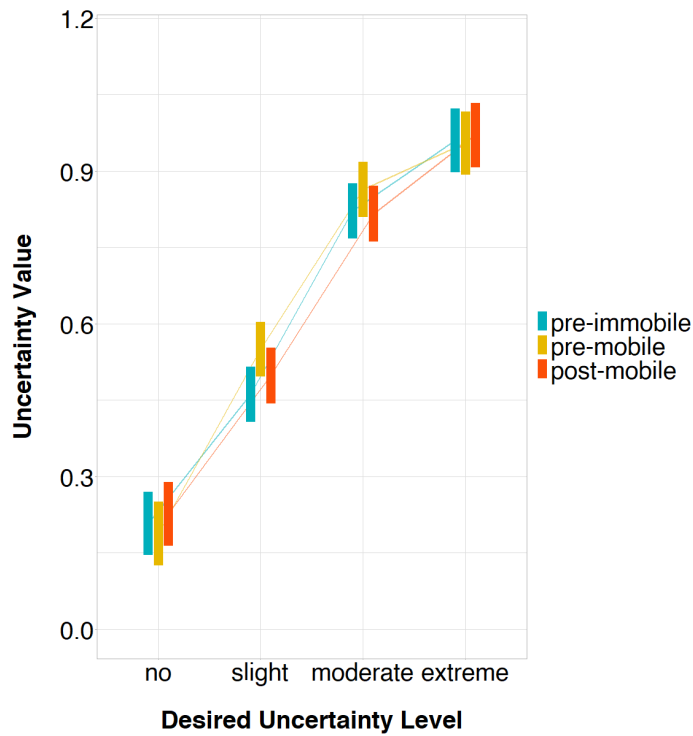

**Feasibility and Usability of the RCUA**

Participant responses on the spatial navigation skills instruments are summarized in Table 6. In regard to participants' real-time use of the RCUA joystick: 13% (n = 9) self-reported that they never forgot to use the device to report wayfinding Uncertainty; 49% (n = 34) self-reported they forgot to use the device one or two times; 19% (n = 13) self-reported that they forgot to use the device 3–5 times; 13% (n = 9) self-reported they forgot to use the device 6–10 times; and 6% (n = 4) self-reported that they forgot to use the device more than 10 times. Since the average number of Uncertainty changing events during each wayfinding task was 25.7, this level of non-reporting was within an acceptable range (approximately 4.08% overall).

**Table 6**

*Individual Differences among the Study Participants*

|  | Scale | Minimum | Maximum | Mean | SD |
|---|---|---|---|---|---|
| All | Mental Rotation (n = 53) | 0 | 13 | 6.09 | 3.57 |
|  | Sense of Direction (n = 51) | 2.4 | 6.4 | 4.33 | 0.98 |
|  | Spatial Anxiety (n=52) | 1 | 4.75 | 2.55 | 0.85 |
| Novice | Mental Rotation (n = 42) | 0 | 13 | 5.95 | 3.57 |
|  | Sense of Direction (n = 41) | 2.4 | 6.4 | 4.26 | 1.03 |
|  | Spatial Anxiety (n = 42) | 1 | 4.75 | 2.56 | 0.91 |
| Expert | Mental Rotation (n = 11) | 2 | 13 | 6.64 | 3.72 |
|  | Sense of Direction (n = 10) | 3.2 | 5.8 | 4.63 | 0.72 |
|  | Spatial Anxiety (n = 10) | 2 | 3.88 | 2.54 | 0.58 |

*Note: Scale ranges were as follows: Mental Rotation 0–24; Sense of Direction 1–7; Spatial Anxiety 1–5.*

Regarding the task load scores (NASA-TLX), participants reported a moderate level of mental task load ($M$ = 4.87, $SD$ = 2.58, 95% CI [4.17, 5.57]) and a low level of physical task load ($M$ = 2.70, $SD$ = 1.73, 95% CI [2.23, 3.18]). One notable result on this instrument is that participants also reported a high perception of success in using the device ($M$ = 7.48, $SD$ = 2.19, 95% CI [6.88, 8.08]) (Table 7). Our survey question regarding the interference of RCUA with wayfinding activities found a relatively low interference level (*Possible Range* 1–10, $M$ = 3.11, $SD$ = 2.04, 95% CI [2.55, 3.67]). Overall, these results indicate that individuals were able to use the joystick device during real-world wayfinding tasks without excessive workload or distraction.

**Table 7**

*Self-reported Task Load (NASA-TLX Scores) for the Use of RCUA during Wayfinding*

|  | **Mean** | **SD** | **95% CI** |
|---|---|---|---|
| Mental Demand | 4.87 | 2.58 | [4.17, 5.57] |
| Physical Demand | 2.70 | 1.73 | [2.23, 3.18] |
| Temporal Demand | 3.17 | 2.29 | [2.54, 3.79] |
| Performance | 7.48 | 2.19 | [6.88, 8.08] |
| Effort | 4.43 | 2.06 | [3.86, 4.99] |
| Frustration | 3.52 | 2.12 | [2.94, 4.10] |

*Note: All metrics in the table range from 1 to 10.*

## Discussion

This methodological research was conducted to help answer the call for better empirical study on human wayfinding processes, and specifically on the role of Uncertainty during spatial navigation (Keller et al., 2020). We evaluated two different approaches to continuous Uncertainty annotation during wayfinding, one in real-time (RCUA) and another using post-wayfinding video evaluation (CUA). The results of the study support the reliability and validity of both measurement approaches. The usability testing for the RCUA also confirmed that our participant sample, which notably included a large representation of adults aged 60 years and older, were able to use the joystick device during real-world wayfinding tasks without excessive workload or distraction.

**Tools' Reliability and Validity**

RCUA approach does not allow participants to see visual feedback displaying their current annotation input level, as is presented on-screen during the CUA annotation. We were concerned that this might lead to reliability problems for the RCUA if participants could not accurately and consistently use the device to report the Uncertainty values that they intended to report. The reliability results (Table 4) showed this concern to be unfounded, as participants of all ages demonstrated great success in consistently using the RCUA joystick to report a specified Uncertainty level.

The validity testing also yielded positive results for both instruments. In terms of convergent validity, the Uncertainty measures collected via the RCUA and the CUA were strongly correlated with participants' reporting of Uncertainty on a written Likert-scale instrument. Discriminative validity was evaluated by comparing the scores of Novice vs. Trained wayfinders and Novice vs. Expert wayfinders. The results indicated that both the RCUA and the CUA were able to distinguish between these groups, though they both had greater success distinguishing Novice vs. Trained than distinguishing Novice vs. Expert. This result is not particularly surprising, since in our group categories the "Trained" wayfinders received relevant route information immediately prior to the wayfinding tasks, while the "Expert" wayfinders simply had prior overall experience with the buildings. Thus, it is plausible to find that Trained group showed greater differentiation from the Novices than did the Experts. Finally, the results indicated that both the RCUA and the CUA had good predictive validity, as participants' scores on these instruments were able to predict their wayfinding performance in terms of task Completion Time and the total Distance Travelled while finding a destination. It is a reasonable assumption that this performance relationship is due to increased Uncertainty being associated with an increase in pausing, backtracking, and information-seeking behaviors (Hirsch et al., 2012), which would increase the task duration and distance covered in the wayfinding tasks.

**RCUA vs. CUA**

The two measurement approaches that were evaluated in this study have different potential strengths and weaknesses. Though we did not find any significant differences between the RCUA and CUA uncertainty metrics for varied wayfinding tasks, the fine-grained analysis of sign seeing events (hypothesis 3) demonstrated that RCUA better captured fleeting real-time uncertainty experience compared with CUA. The result showed that multi-tasking and a lack of visual feedback didn't prevent most participants from using RCUA to report uncertain experiences in a timely manner. On the contrary, CUA measure omitted details because it relied on participants' memory and analysis of their prior activities. As for the amplitude between two methods, though we didn't find significant difference at the task level, sign seeing analysis showed that participants tended to report less uncertain experience using CUA. It could result from systematic underestimation due to memory bias and annotation setting, or participants overly reported uncertain experience using RCUA, which requires further examination. Combined with the finding of strong discriminative and predictive validity, both instruments can serve as effective approaches for measuring wayfinding Uncertainty at the task-level resolution. However, RCUA better captured fine-grain dynamics of uncertain experience.

It is worth noting in this regard that the average time from the participants' completion of the wayfinding tasks to their completion of the CUA video-based reporting was about 37 minutes ($M = 36.72$, $SD = 9.94$, range: 18.55–67.22). It is possible that longer time intervals could reduce the accuracy of the CUA approach. Another vital component of the CUA is that participants were annotating videos of their own previous wayfinding activities (not a video of another person); the results of the current study should not be generalized to support video-based annotation made by observers.

**Figure 8**

*One Application of the RCUA: Kernel Density Map Showing Uncertainty Distribution from All Participants of One Floor Level*

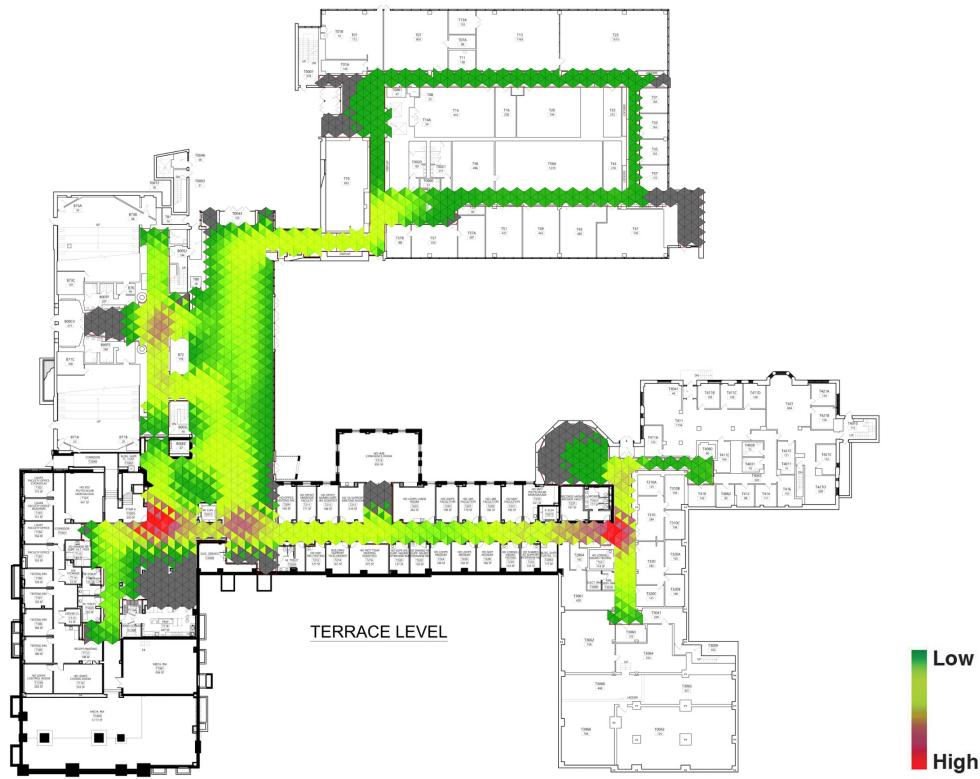

## Impact

The methods developed and evaluated here for wayfinding Uncertainty annotation can enable the collection of fine-grained, temporal data in real-world environments. In many wayfinding research applications this approach can improve upon the traditional discrete metrics such as task completion times, distance covered, or number of mistakes. For example, when evaluating wayfinding design strategies or considering renovations in complex buildings such as a hospital or an airport, continuous wayfinding annotation can enable rigorous scientific studies to compare the efficacy of different interior designs or signage strategies, or to locate problem spots within a building floorplan (Figure 8). In the broader trajectory of the field, this can contribute significantly to the development of general knowledge about wayfinding behaviors and effective designs.

Another valuable potential of studies that apply the RCUA or CUA approach is that they can help to highlight individual differences in reactions to wayfinding Uncertainty and the associated behavioral changes (e.g., different ways in which spatial information is sought). Such insights, which cannot be obtained by studying discrete performance metrics, may further expand our knowledge of effective wayfinding designs that are suitable for diverse users. Designers and architects are limited by the extent of the prior experience and imagination, which can sometimes lead them to overlook the needs and reactions of various individuals, such as people with colorblindness, illiteracy, or various forms of neurodivergence. By studying continuous wayfinding behaviors among diverse participants, researchers can better identify and respond to such needs. In addition to improving interior designs, such research may also provide the grounding for developing effective smart navigational aid systems that use physiological signals to classify human uncertainty states. In high-stakes situations, such as those involving emergency first responders or helping patients to reach the appropriate care centers, providing effective and appropriately intrusive information when wayfinding Uncertainty arises can be tremendously important.

Interest in testing novel hypotheses about the impact of environmental features on human wayfinding behavior spans across numerous architectural specialties, as well as other "spatial professions" such as the design of immersive virtual environments, video games, and spherical cinema (Gepshtein & Snider, 2019; Darfler et al., 2022, Kalantari et al., 2022). It also extends to related fields such as environmental psychology and neuroscience. Thus, it is our aspiration that the validated RCUA and CUA approaches will provide such researchers with a valuable tool. The hardware components of the RCUA and CUA systems are relatively low-cost and accessible, and they can be combined either with our freely available software components or with new researcher-created tools. The continuous data-collection approach for annotating Uncertainty states is also well-suited for integration with physiological measures such as EEG and gaze-tracking, which can further enhance our knowledge of human wayfinding behavior and cognition.

**Limitation and Future Directions**

Given the challenges in designing a mobile, self-reported continuous measurement system, there were naturally some limitations to our work. First, it should be noted that the temporal accuracy of the RCUA and CUA has not yet been validated. We plan to conduct a future study to ascertain how well these annotations systems capture the time-period in which Uncertainty occurs, and to see if any adjustments need to be made to account for lag or to improve the temporal accuracy. Second, more work needs to be done to determine if the RCUA approach is suitable for research that requires other types of simultaneous participant action, such as reporting additional continuous variables, holding a map, or using a VR controller. The more multi-tasking that an experiment involves, the more likely it is that the effectiveness of real-time self-reporting may suffer. The validation results of the current study should not be extended to such contexts that place additional burdens on participants' attention.

The design of our experiment procedure also gives rise to some cautions and limitations. The participants in the wayfinding study were followed by a researcher who carried various position-tracking and data-logging equipment. While the researchers tried to remain at least 5 ft.

behind the participant at all times, their presence may have affected the Uncertainty behaviors of the participants, especially in regard to remembering to use the reporting joystick. Additional research is needed to determine if RCUA is effective during solo-wayfinding experiments. Another possible concern is that our evaluation of convergent validity was somewhat simplistic, relying on a single Likert-scale response after each wayfinding task to compare the RCUA and CUA data. The ability to better establish convergent validity is a bit problematic, as to the best of our knowledge there are currently no other scales measuring this continuous construct. However, the evaluation could be made more effective by using a more complex Likert survey and/or correlating Uncertainty with related types of annotation such as "feeling lost." Finally, we restricted the context of the study to in a large indoor educational setting, and did not test the Uncertainty reporting in outdoor environments or other building types. We do not expect that such contexts will have a pronounced effect on the use of the CUA, but it is possible that environments with greater crowding or other distractions could negatively impact the real-time reporting of RCUA data.

## Conclusion

In this study we developed and evaluated two novel methods to continuously measure self-reported wayfinding Uncertainty levels. We confirmed the methods' usability and reliability, as well as their convergent, discriminative, and predictive validity. Both the real-time (RCUA) and video-based (CUA) approaches were shown to be effective means of identifying the extent of Uncertainty that participants were experiencing throughout the wayfinding tasks.